\documentclass[aps,pra,twocolumn,superscriptaddress]{revtex4-1}

\usepackage{graphicx}
\usepackage{dcolumn}
\usepackage{bm}
\usepackage{amssymb}
\usepackage{amsmath}
\usepackage{subfigure}
\usepackage{epstopdf}
\usepackage{url}
\usepackage[breaklinks]{hyperref}
\usepackage{breakurl}

\newif\ifhyper
\hypertrue
\ifhyper
\hypersetup{
   citecolor = {green},
   colorlinks = {true},
   urlcolor = {blue}
}
\fi

\usepackage{color}
\usepackage{soul}

\begin{document}
\newpage
\title{Implementation of a Hybrid Classical-Quantum Annealing Algorithm For Logistic Network Design}

\author{Yongcheng Ding}
\email{jonzen.ding@gmail.com}
\affiliation{International Center of Quantum Artificial Intelligence for Science and Technology (QuArtist) \\ and Department of Physics, Shanghai University, 200444 Shanghai, China}
\affiliation{Department of Physical Chemistry, University of the Basque Country UPV/EHU, Apartado 644, 48080 Bilbao, Spain}
\author{Xi Chen}
\email{xchen@shu.edu.cn}
\affiliation{International Center of Quantum Artificial Intelligence for Science and Technology (QuArtist) \\ and Department of Physics, Shanghai University, 200444 Shanghai, China}
\affiliation{Department of Physical Chemistry, University of the Basque Country UPV/EHU, Apartado 644, 48080 Bilbao, Spain}
\author{Lucas Lamata}
\affiliation{Department of Physical Chemistry, University of the Basque Country UPV/EHU, Apartado 644, 48080 Bilbao, Spain}
\affiliation{Departamento de F\'isica At\'omica, Molecular y Nuclear, Universidad de Sevilla, 41080 Sevilla, Spain}
\author{Enrique Solano}
\email{enr.solano@gmail.com}
\affiliation{International Center of Quantum Artificial Intelligence for Science and Technology (QuArtist) \\ and Department of Physics, Shanghai University, 200444 Shanghai, China}\affiliation{Department of Physical Chemistry, University of the Basque Country UPV/EHU, Apartado 644, 48080 Bilbao, Spain}
\affiliation{IKERBASQUE, Basque Foundation for Science, Plaza Euskadi 5, 48009, Bilbao, Spain}
\affiliation{IQM, Nymphenburgerstr. 86, 80636 Munich, Germany}
\author{Mikel Sanz}
\email{mikel.sanz@ehu.eus}
\affiliation{Department of Physical Chemistry, University of the Basque Country UPV/EHU, Apartado 644, 48080 Bilbao, Spain}
\affiliation{IKERBASQUE, Basque Foundation for Science, Plaza Euskadi 5, 48009, Bilbao, Spain}
\affiliation{IQM, Nymphenburgerstr. 86, 80636 Munich, Germany}

\date{\today}

\begin{abstract}
The logistic network design is an abstract optimization problem that, under the assumption of minimal cost, seeks the optimal configuration of the supply chain's infrastructures and facilities based on customer demand. Key economic decisions are taken about the location, number, and size of manufacturing facilities and warehouses based on the optimal solution. Therefore, improvements in the methods to address this question, which is known to be in the NP-hard complexity class, would have relevant financial consequences. Here, we implement in the D-Wave quantum annealer a hybrid classical-quantum annealing algorithm. The cost function with constraints is translated to a spin Hamiltonian, whose ground state encodes the searched result. As a benchmark, we measure the accuracy of results for a set of paradigmatic problems against the optimal published solutions (the error is on average below $1\%$), and the performance is compared against the classical algorithm, showing a remarkable reduction in the number of iterations. This work shows that state-of-the-art quantum annealers may codify and solve relevant supply-chain problems even still far from useful quantum supremacy.
\end{abstract}

\maketitle
\section{Introduction}
Calculating the global minimum (maximum) of a multi-variable function is in general arduous, especially when the number of variables and constraint conditions grows and the objective function is highly non-linear. The problem is in the NP-hard complexity class since it is complicated even to verify whether a given solution is the optimal one. As a branch of the optimization problem, the logistic network design problem (NDP) covers a massive set of decision-making problems for management issue~\cite{ballou}, e.g., where to place facilities and how to assign customers minimizing the total cost, or how to redistribute driving paths of vehicles to reduce traffic jams. There is an endless list of classical algorithms for optimization problems, e.g., branch and bound, context partition and dynamical programming, metaheuristic algorithm based on a single solution such as hill-climbing, simulated annealing~\cite{sa}, and tabu search~\cite{tabu1,tabu2}, or intelligent optimization by genetic algorithms~\cite{ga}, ant colony optimization~\cite{aco1,aco2}, and artificial neural networks~\cite{ann}. This algorithm has been applied to study NDP and provided some preliminary results~\cite{prelim1,prelim2,prelim3,prelim4}. However, these algorithms' mathematical principles for finding global minima are not systematically established and, in most cases, it requires experience adjusting parameters. This raises the demand for developing an interpretable algorithm for solving logistic NDPs efficiently.

Quantum annealing is an optimization technique especially suitable for satisfiability problems, which makes use of quantum tunneling of potential barriers to enhance the performance of the classical algorithm~\cite{annealing1,annealing2}. The ground state codifying the solution of the problem is attained expectedly employing a shorter annealing time than the classical algorithm and afterward decoded to achieve the optimal solution with respect to the objective function~\cite{nielsenchuang}. Current D-Wave cloud quantum annealer comprises 2048 qubits distributed in a hardware architecture according to the Chimera graph. The constraints imposed by the architecture generally allow for codifying only relatively small problems, which can be enhanced when combined with classical algorithms. Additionally, this quantum device is affected by thermal fluctuations, decoherence, and I/O errors, which reduces the signal-to-noise ratio and consequently prolongs the computation time due to the extra sampling required for canceling the noise (otherwise, the accuracy of the solution would be affected). Nonetheless, quantum annealers have proven their capability to codify hard problems in different fields, such as condensed matter physics~\cite{cmp-spinglass-theory,cmp-manufactured,cmp-traverse,cmp-disordered,cmp-experiment}, engineering~\cite{volkswagen}, cryptography~\cite{FengHu,FengHu2}, biology~\cite{bio} and finance~\cite{quantfin1,quantfin2,quantfin3}, among others. This shows that current quantum annealing technology, although incoherent and still far from reaching useful quantum advantage, is ready to implement relevant small-scale optimization problems.

In this Article, we experimentally address a set of paradigmatic logistic NDPs employing a hybrid classical-quantum annealing algorithm, showing a remarkable accuracy (less than 1\% error) despite the device incoherence and better performance in the number of iterations with respect to the classical one (about one-tenth, when classical annealing algorithm can reach global minimum with adequate hyperparameters). Inspired in Ref.~\cite{CSA}, we propose an alternative approach, which we call {\it combined quantum annealing algorithm}, that makes use of two layers with feedback-control interaction between them. The approach is tested in $12$ paradigmatic logistic NDPs employing both the simulator and the D-Wave cloud quantum annealer, achieving the aforementioned remarkable results when compared with the classical ones. This supports the extended idea that a hybrid quantum-classical algorithm will allow us to enlarge the class of solvable problems with quantum computers (quantum annealers), accelerating the development of quantum information processing tasks in quantum technologies.

This manuscript is organized as follows: in Sec.~\ref{sec:model}, we formulate the logistic NDP as a constrained $0-1$ programming problem, which will afterwards allow us to map it into D-Wave's Chimera architecture. In Sec.~\ref{sec:CQSA}, we introduce the fundamental elements of quantum annealing and of the combined quantum annealing algorithm. Afterwards, in Sec.~\ref{sec:exp}, we experimentally tests the optimization of logistic NDPs by comparing the results against the best known classical ones, as well as those given by a classical algorithm for finding global minimum. Finally, in Sec.~\ref{sec:discussion}, the results are analyzed and possible alternative software and hardware approaches for further enhancement are discussed. The conclusions are finally listed in Sec.~\ref{sec:conclusion}.

\section{Logistic Model\label{sec:model}}
Although a logistic NDP could be described in an abstract framework, we choose the customer-facility picture because it is illustrative. Let us assume that there are at most $m$ sites for potential facilities and $n$ customers to be allocated. The indices $J=\{1,2,\cdots,m\}$ and $I=\{1,2,\cdots,n\}$ denote the set of potential location sites and the set of customers, respectively. It costs $f_j$ to build a facility with production capacity $v_j$ placed at site $j \in J$. When a customer $i\in I$ with product demand $d_i$ is served by facility $j$, the transportation process brings $c_{ij}$ to the cost. One has to find a way to allocate customers with a minimum total cost that ensures all customers are served, and none of the facilities is overflowed. To formulate the model, one has to minimize the following cost function
\begin{equation}
\label{eq:cost-function}
\text{cost}(x_j,y_{ij})=\sum_j f_jx_j+\sum_i\sum_j c_{ij}y_{ij},
\end{equation}
where $x_j$ and $y_{ij}$ are binary variables that represent the allocation configuration. A facility is built at site $j$ when $x_j=1$, and obviously, customers can only be served from sites with facilities. Accordingly, $y_{ij}=1$ means customer $i$ is assigned to the facility in site $j$, and other facilities are no longer available for this customer. These constraints can be expressed as
\begin{equation}
\label{eq:equ}
\sum_j y_{ij}=1,\ \forall i\in I,
\end{equation}
\begin{equation}
\label{eq:inequ}
\sum_i d_iy_{ij}<v_j,\ \forall j\in J,
\end{equation}
\begin{equation}
\label{eq:legit-position}
y_{ij}\leq x_j,\ \forall i\in I,\ \forall j\in J.
\end{equation}
Minimizing the cost function under the constraints above is proven to be an NP-hard problem, i.e., obtaining its global minimum value is computationally challenging and highly time-consuming. Additionally, it is also hard to verify if a given network is the solution that minimizes the cost function. These features lead to the demand for specific algorithms that accelerate the searching process and enhance the solution's quality.

\section{Combined Quantum Annealing Algorithm\label{sec:CQSA}}
The solution of satisfiability problems can be codified in the ground state of a problem spin Hamiltonian $H_\text{pro}$ \cite{FGGS00}. Quantum annealing is a quantum algorithm based on the adiabatic theorem \cite{AE99}, which ensures that if we start in the ground state of a known Hamiltonian $H_0$, by slowly modifying a parameter $t$ transforming $H_0$ into $H_\text{pro}$, the system remains in its ground state, providing us with the solution for the satisfiability problem. For example, in a spin-$1/2$ annealer, the total Hamiltonian is split into a tunneling Hamiltonian and a problem Hamiltonian, which codifies the solution for the problem,
\begin{equation*}
H=\left(1-\frac{t}{t_f}\right)\sum_i \hat{\sigma}^x_i+\frac{t}{t_f}\left(\sum_i h_i\hat{\sigma}^z_i+\sum_{i>j}J_{ij}\hat{\sigma}^z_i\hat{\sigma}^z_j\right).
\end{equation*}
The system is supposed to be in the ground state of the problem Hamiltonian when $t=t_f$. If we introduce the qubit operator $\hat{x}$ with eigenvalues $0$ and $1$, such that $\hat{x}|0\rangle=|0\rangle$ and $\hat{x}|1\rangle|1\rangle$, respectively, {\it quadric unconstrained binary optimization} (QUBO) problem can be mapped to a spin-1/2 quantum annealing problem by $\hat{x}=(1+\hat{\sigma}^z)/2$.

Looking at the cost function given by Eq.~\eqref{eq:cost-function}, we understand that it cannot be directly optimized by quantum annealing since there are constraints. The cost function and all constraints are linear, so they are all relatively easy to be converted to a QUBO formulation. For example, Eq.~\eqref{eq:legit-position} can be satisfied by introducing penalty terms in the form $\alpha(x_j-y_{ij}-z_{ij})^2$, where $z_{ij}$ are auxiliary binary variables. One can verify that the penalty is unavoidable only when $x_j=0$ and $y_{ij}=1$. With an effective QUBO formulation satisfying Eq.~\eqref{eq:equ} and Eq.~\eqref{eq:inequ}, we propose a hybrid quantum-classical algorithm inspired in Ref.~\cite{CSA} for obtaining the global (or quasi-global) minimum. We apply a combined quantum annealing algorithm comprising two layers, namely, the outer and the inner layers. In the former, a simulated annealing algorithm performs the optimization for facilities' locations while, in the latter, the quantum annealing process runs.

In the outer layer, a list with elements $x_j$, sorted by index $j$ in ascending order, denotes the neighboring configuration of facilities, while a neighboring function operates on the list to generate a new configuration in three ways: (\romannumeral1) pick a $x_j$ with value $1$, and set it to $0$, which means a facility is randomly closed; (\romannumeral2) pick a $x_j$ with value $0$, and set it to $1$, which means a facility is randomly built on a site; (\romannumeral3) swap the values of $x_{j}$ and $x_{k}$, if their values are different, which means a facility moves to another site. Dice will be rolled to decide which operation is carried out by the neighboring function on the list, while all the operations should be allowed, e.g., when all facilities are open, operation (\romannumeral2) is no longer available.
\begin{figure}
\includegraphics[scale=0.5]{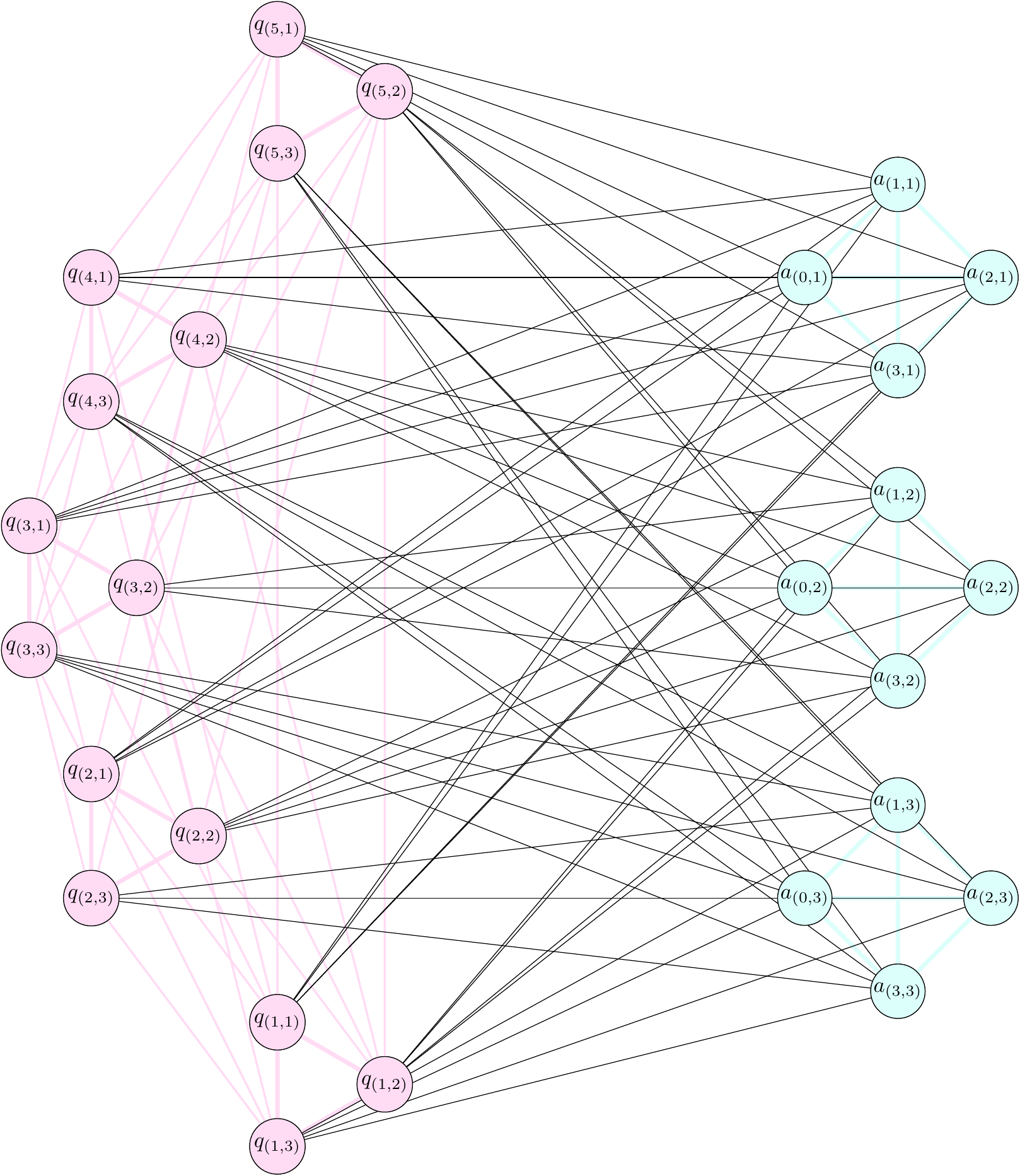}
\caption{\label{fig:scheme} A scheme of how binary variables and constraint conditions are encoded by logical and ancilla qubits. Here, we represent a simple NDP problem comprising $3$ facilities and $5$ customers, as an illustrative example, while different colors and thicknesses classify the couplers according to their properties.}
\end{figure}
In the inner layer, we perform quantum annealing to minimize $\sum_i\sum_j c_{ij}y_{ij}$ under constraint given by Eq.~\eqref{eq:equ}, Eq.~\eqref{eq:inequ}, and Eq.~\eqref{eq:legit-position}. As we mentioned before, the latest quantum annealer is designed to solve QUBO problems, i.e. to minimize the objective function $\text{obj}(\bf{x})=\bf{x}^{\text{T}}Q\bf{x}$, that $Q$ is the QUBO matrix and $\bf{x}$ is the binary vector. We map binary variables $y_{ij}$ to qubit $q_{(i,j)}$, and allowed $j$ are the sites on which facilities are built. In order to introduce the constraint given by Eq.~\eqref{eq:equ}, we employ weighted penalty functions, with a reasonable $\lambda_i$, yielding an effective unconstrained Hamiltonian for customer $i$, e.g. $\lambda_i(\sum_j q_{(i,j)}-1)^2$, which guarantees that customer $i$ is associated only to one facility. Accordingly, Eq.~\eqref{eq:inequ} may also be written as ${\mu_j(\sum_i d_i q_{(i,j)}+\langle\bf{2},\bf{a}_j\rangle}-v_j)^2$ for facility $j$, with slack variables encoded by ancilla qubits. Here, $\langle\bf{2},\bf{a}_j\rangle$ denotes the binary expansion $\sum_{l=0}^k 2^la_{(l,j)}$, which implies that the number of ancilla qubits required to introduce the constraint is $k_j=\lceil\log_2 v_j\rceil$. One can verify that any violation of Eq.~\eqref{eq:equ} and~\eqref{eq:inequ} introduce extra squared cost, being scaled by $\lambda_i$ and $\mu_i$. For example, if customer $i$ is not assigned to any facility, i.e., $\sum_j q_{(i,j)}=0$, the violation of Eq.~\eqref{eq:equ} punishes a cost of  $\lambda_i$, which should be larger than the profit from saving the transportation cost $c_{ij}$. The problem Hamiltonian is hence given by
\begin{eqnarray}
\label{eq:problem}
H_P&=&\sum_i\sum_j c_{ij}q_{(i,j)}+\sum_i{\lambda_i\Big(\sum_j q_{(i,j)}-1\Big)^2} \nonumber\\
&+&\sum_j{\mu_j \Big(\sum_i d_i q_{(i,j)}+\langle\bf{2},\bf{a}_j\rangle}-v_j \Big)^2,
\end{eqnarray}
which could be mapped to a solvable spin-$1/2$ Hamiltonian for the annealer. The ground state of the problem Hamiltonian is supposed to be the configuration that minimize the classical objective function with reasonable penalty coupling strengths $\lambda_i$ and $\mu_j$.

This combined quantum annealing algorithm works as follows: (\romannumeral1) Set the optimal cost to infinity and generate an initial configuration as the optimal configuration of facilities. Schedule the annealing process, i.e., the cooling rate, initial temperature, final temperature, etc. (\romannumeral2) Flip the values of variables $x_j$ in list by neighboring function and obtain a new configuration $\tilde{x}_j$. (\romannumeral3) Perform the quantum annealing process according to the neighboring configuration, decode the state of qubits to $\tilde{y}_{ij}$, and calculate value of the cost function $\text{cost}(\tilde{x}_j,\tilde{y}_{ij})$. (\romannumeral4) Apply Metropolis algorithm such that, if $\text{cost}(x_j,y_{ij})>\text{cost}(\tilde{x}_j,\tilde{y}_{ij})$, we accept the new configuration and value as the optimal ones and go for next step. Otherwise, randomize $\rho\in(0,1)$, if $\rho<\exp(-(\text{cost}(\tilde{x}_j,\tilde{y}_{ij})-\text{cost}(x_j,y_{ij}))/T)$, we also accept them and continue. We go for the next step without operations if the new configuration and value are denied. (\romannumeral5) Increase the iteration index by one and check if it meets the upper limit. Once it is larger than the maximum iteration number, we adjust the annealing temperature according to the schedule, reset the iteration index, and go on for the next step. Otherwise, return to step (\romannumeral2). (\romannumeral6) Output the optimal solution if the current temperature is not higher than the target temperature. Otherwise, return to step (\romannumeral2).

Thus, if all parameters in the outer and inner layers were correct and the quantum annealer was noiseless, the places for building facilities, the allocation of customers, and the optimized total cost would be obtained by this combined annealing algorithm. In any case, error correction can be applied to provide a quasi-optimal solution, which is a relevant advantage since we are employing a noisy and incoherent quantum annealer.

\section{Experiments\label{sec:exp}}
We choose the same twelve problems tested by Ref.~\cite{CSA}, which are open-source NDP test problems from OR-Library~\cite{OR-Library}. The optimal solution is given by the author using \textit{Lindo} software. We encoded $m\times n$ logical qubits by $q_{(i,j)}$ and $\sum k_j$ ancilla qubits by $a_{(l,j)}$ for NDP problem with $m$ potential sites for building facilities with capacities $v_j$ and $n$ customers to be served. The connectivity of these qubits is high, with a number of couplers approximately given by
\begin{equation}
\frac{(m-1)mn}{2}+\frac{(n-1)nm}{2}+\sum_j\frac{(k_j-1)kn}{2}+\sum_j nk_jm,
\end{equation}
where terms denote the numbers of couplers between logical qubits given by Eq.~\eqref{eq:equ}, Eq.~\eqref{eq:inequ}, couplers between ancilla qubits and couplers that link an ancilla qubit to a logical qubit (see Fig.~\ref{fig:scheme}), respectively.
The minimal number of qubits for solving this problem is $n\times m+k\times m$, if the structure of the quantum device is exactly the same as we showed in Fig.~\ref{fig:scheme}. Notice that this number could substantially grow if qubits are connected differently from the problem's connectivity graph since additional ancillary qubits are required to implement minor embedding. The smallest problem for testing requires 1056 qubits with 40320 couplers, which cannot be directly implemented in the 2048-qubit D-Wave quantum annealer due to its limited connectivity (6016 couplers). The open-source software \textit{qbsolv} developed by D-Wave \cite{qbsolv} allows for splitting a large QUBO problem into smaller embeddable sub-problems, which can subsequently be solved in either a local simulator with tabu algorithm or the real quantum annealer under authorized license. The QUBO problem is submitted to a server for queueing, and the result is retrieved after the annealing is performed. Although the computing time in a real annealer is negligible, one trial of solving a large QUBO matrix could be very time-consuming (about 15 minutes for a $900\times900$ QUBO matrix) due to the queuing time. We also emphasize that $x_j$ vanishes since we employ a hybrid algorithm instead of a standard approach. In the inner layer where quantum annealing performs, we solve a subproblem as follows: how customers should be assigned according to a given facility configuration? Some variables $y_{ij}$ are constantly zero if there is no facility built at site $x_j$. In our practice, these variables are already removed in each quantum annealing process, making embedding or solving by \textit{qbsolv} easier and more efficient. Thus, in each iteration that generates a new facility configuration, the number of logical qubits required is $\tilde{m}\times n$ with a maximum of $m\times n$, where $\tilde{m}$ denotes the number of facilities being built.

The initial temperature of simulated annealing is set to be 10000, which is scaled to a reasonable value considering the deviation between the new and old value of the cost function. For simplicity, we choose $\alpha=0.5$ as cooling rate and the schedule as scale cooling. The maximum iteration number is $m$, and the target temperature is 1. To obtain the optimal solution according to a given configuration of facilities, the penalty strength should be sufficiently strong, i.e., it should ensure that the problem Hamiltonian's ground state satisfies all constraint conditions. Hence, if the quantum annealer were noiseless, the parameters would follow $\lambda_j\gg\mu_i\gg c_{ij}$, which ensures that every customer is assigned to only one facility, none of the facility is overflowed, and the configuration corresponds to the minimum total cost. However, in practice, the quantum device is affected by thermal fluctuation,  inaccuracy of magnetic field tunneling, and energy excitation by nonadiabatic effects. Thus the result in general satisfies the constraints, but the solution is not optimal. Consequently, the penalty strength is set slightly larger than $c_{ij}$ for the possibility of obtaining an optimal solution. Although sometimes constraints are also not fulfilled due to the noise, this could be corrected by repeating the quantum annealing process until every constraint is satisfied. After studying the dataset, the penalty strength $\lambda_i$ are set to be $\min_j(c_{ij})$, while $\mu_i$ are almost negligible considering the demand and capacities.
\begin{table}
\begin{ruledtabular}
\begin{tabular}{cccccc}
Problem&Size&Lindo&SA&QA\\
\hline
cap71&$16\times50$&932615.7500&1460909.750&933172.1000\\
cap72&$16\times50$&977799.4000&1395389.538&977988.1000\\
cap73&$16\times50$&1010641.450&1585875.550&1010641.450\\
cap74&$16\times50$&1034976.975&1390963.787&1034976.975\\
cap101&$25\times50$&796648.4400&1182235.563&797656.2875\\
cap102&$25\times50$&854704.2000&1282306.175&854952.5125\\
cap103&$25\times50$&893782.1125&1395701.200&894872.1125\\
cap104&$25\times50$&928941.7500&1458550.450&928941.7500\\
cap131&$50\times50$&793439.5620&1167543.950&796066.6500\\
cap132&$50\times50$&851495.3250&1132436.300&852291.9375\\
cap133&$50\times50$&893076.7120&1126423.238&893521.4125\\
cap134&$50\times50$&928941.7500&1321380.713&928941.7500\\
\end{tabular}
\end{ruledtabular}
\caption{\label{table:result}The combined quantum annealing algorithm is tested by 12 NDP problems from OR-Library while the results are listed under QA. The optimal solutions are compared with the global optimal solutions from \textit{Lindo} given by the author of Ref.~\cite{CSA} and classical simulated annealing algorithm with the same annealing schedules and iteration numbers denoted by SA. In our experiments, the combined quantum annealing algorithm outperforms simulated annealing as a well-known metaheuristic algorithm massively under the same iteration numbers.}
\end{table}

\begin{figure*}
\includegraphics[scale=0.45]{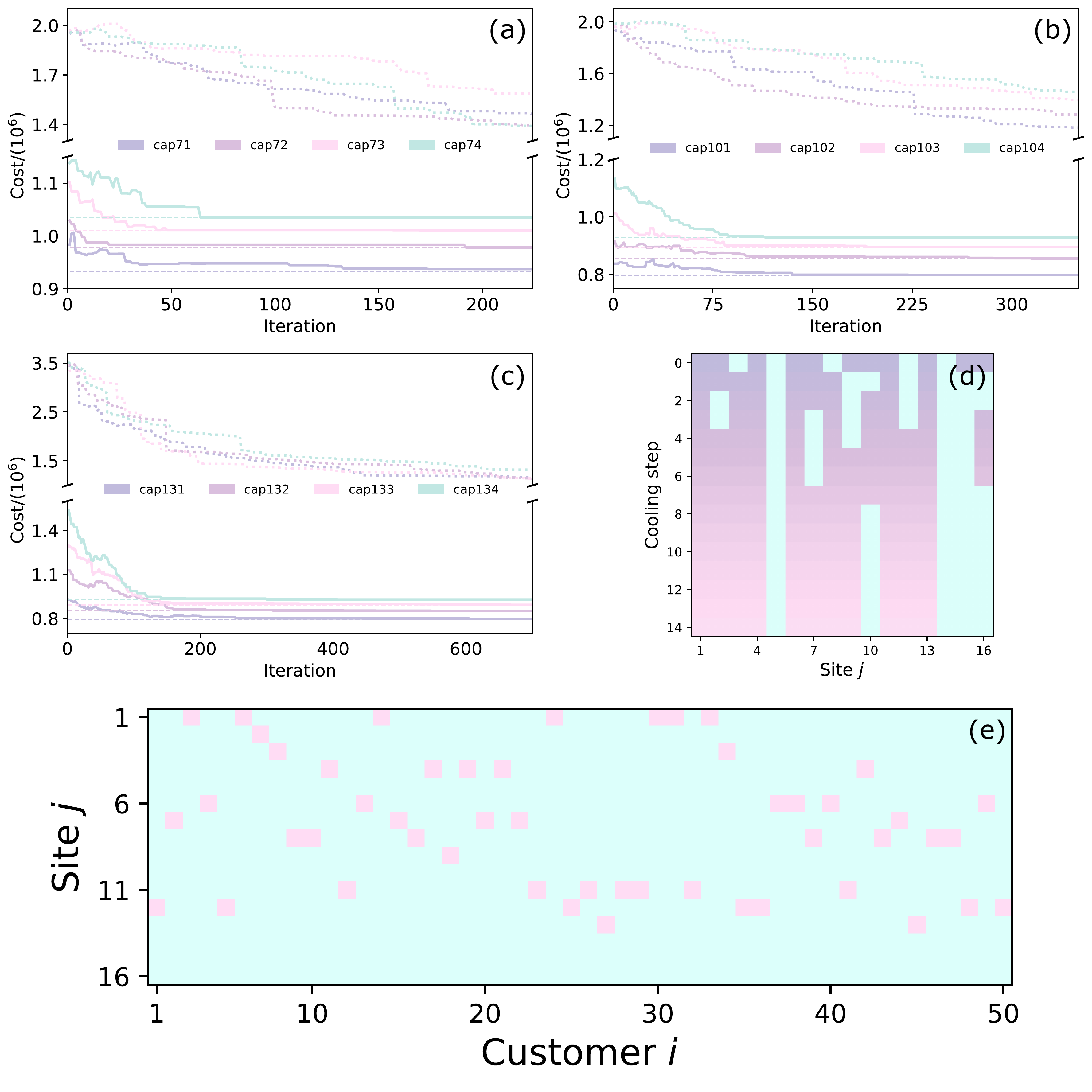}
\caption{\label{fig:result}(a-c) Optimal costs versus iteration numbers, with different problem sizes. Different problems with the same size are in the same color. The evolutions of the combined annealing algorithm and simulated annealing algorithm are plotted by lines and dotted lines, with ground energies by dashed lines for comparison. (d,e) Facility configuration at each cooling step and the allocation of customers of Problem cap71 shown in Table.~\ref{table:result}, with a minimum cost of 933172.1000. A block in warm/cold color denotes $x_j,y_{ij}=1$ or $0$.}
\end{figure*}

The results are presented in Table.~\ref{table:result}, while the hyper-parameters for simulated annealing algorithm are the same as those in outer layer of our quantum-classical hybrid algorithm. For a fair comparison, we take $m$ iterations in each cooling step (the simulated annealing algorithm is discussed in Appendix.~\ref{app:SA}). Consequently, both simulated annealing and combined quantum annealing algorithms take $m\times\lceil\log_{0.5}1e-4\rceil=14m$ runs for Metropolis acceptance criterion. Results presented here are not deliberately selected via several trials, which means one can obtain different results fluctuating with a small range. We also show how these two algorithms work for different logistic NDPs by depicting the evolution of optimal costs by iteration steps in Fig.~\ref{fig:result} (a-c). We notice that results given by the classical algorithm are still far from optimal with the same iterations. To obtain a global minimum (or near), one should evaluate about $10m$ new configurations in each cooling step instead of $m$, with no guarantee of success because the classical algorithm highly depends on others' choice hyperparameters.

\section{Discussion\label{sec:discussion}}
Let us analyze the experimental results for improving our understanding of the protocol before any further discussion about how to enhance its performance. As we mentioned above, solving a $900\times900$ QUBO matrix by \textit{qbsolv} takes about 15 minutes, summing up to approximately 56 hours for Problem cap71 (even more for larger problems). This would be solved in case of having local access to the quantum device and, of course, if the QPU would have more qubits and better/customized connectivity for directly running quantum annealing instead of employing \textit{qbsolv}. We highlight that, even though our quantum device is noisy and incoherent, one can still find a result with a remarkable deviation smaller than 0.5\% when compared against the global optimal result obtained from the exhaustive search. We run the quantum annealing process with a large sampling number, selecting the configuration with minimum energy among all samples, even if this configuration might not appear that frequently due to energy fluctuations during the non-adiabatic process. Once the optimal configuration for facilities is obtained, the optimal solution can be checked/refined by repeatedly running the quantum annealing process to check if a customer allocation with a lower cost can be attained. The reason is that the search space by exhaustive searching for a QUBO matrix encoded in $n\times m$ logical qubits and $k\times m$ ancilla qubits is $2^{(n+k)m}$, but the search subspace is dramatically reduced once the quantum annealer excludes most of the states with higher energy. In this way, the quantum annealing algorithm works for searching a global minimum even in noisy intermediate-scale quantum devices. The price to pay is a more extensive sampling in the quantum annealing process.

Now, we evaluate several improvements to make the algorithm more efficient. From the perspective of algorithm, the outer layer applies a classical simulated annealing algorithm which theoretically obtains global minimum with adequate annealing parameters. As we mentioned before, the inappropriate annealing schedule will stuck the algorithm to a local minimum and we should find the initial temperature for simulated annealing. The total number of iterations to guarantee a global minimum is enormous which is proved by Ref.~\cite{hajek}, while the transition probability from state $i$ to state $j$ is denoted by $P_{ij}=G_{ij}A_{ij}$. The acceptance probability follows the Metropolis acceptance criterion that $A_{ij}=\exp(-(E_j-E_i)/T)$ when $E_j>E_i$, otherwise $A_{ij}=1$. We assume that the generation probability is symmetric $G_{ij}=G_{ji}$ and the Markov chain of a given temperature is acyclic and irreducible, then the system follows a Maxwell-Boltzmann distribution that
\begin{equation}
\label{eq:distribution}
\pi_i=\frac{|N(i)|\exp(-E_i/T)}{\sum_j |N(j)|\exp(-E_j/T)},
\end{equation}
where $N(i)$ and $N(j)$ denote the sets of neighbours of state $i$ and $j$, respectively, and generation probability $P_{ij}$ is distributed uniformly among neighbours of state $i$ (i.e., $G_{ij}=1/|N(i)|$ if $j\in N(i)$). Accordingly, the acceptance probability $\chi(T)$ and an iteration algorithm to obtain the proper annealing schedule are provided and proved in Ref.~\cite{walid}. In this way, we could analyze the dataset and find optimal parameters for the outer layer algorithm before solving the NDP problem with combined quantum annealing algorithm. In the inner layer, quantum annealing for a large QUBO matrix cannot be implemented directly which requires \textit{qbsolv} for generating subproblems. However, the partition algorithm in the main loop of \textit{qbsolv} sometimes leads to local minimum that requires better method for splitting the matrix. Alternative algorithms could be introduced, e.g. embedding larger subproblems which exploit the resources provided by the quantum annealer~\cite{improving-qbsolv} or emphasizing the importance of mitigating the embedding cost~\cite{hua}. Meanwhile, the quantum annealing algorithm in the inner layer is affected by the penalty strength, while these parameters are very tricky to be decided. One should scale them according to the dataset and the noise of the hardware, for obtaining an acceptable solution that satisfies all constraint conditions. We notice that the optimized penalty strength for a QUBO problem could be given with the combination of machine learning algorithm, e.g., gradient descent as the most trivial idea, by quantum annealing with variant parameter vectors $\bm{\lambda}$ and $\bm{\mu}$ and stop at an optimal solution. 

As briefly analyzed in Sec.~\ref{sec:CQSA}, the minimization of cost function~\eqref{eq:cost-function} under constraint conditions \eqref{eq:equ},~\eqref{eq:inequ}, and~\eqref{eq:legit-position} can be encoded in QUBO formulation. Specifically, Eq.~\eqref{eq:legit-position} can be implemented in a similar way as penalty terms by using auxiliary qubits $b_{(i,j)}$, for ensuring that customers are not assigned to sites without facility. Thus, one can solve the NDP by standard approach, performing quantum annealing of the problem Hamiltonian, which reads as
\begin{eqnarray}
\label{eq:direct}
H_P&=&\sum_jf_j\tilde{q}_{(j)}+\sum_i\sum_j c_{ij}q_{(i,j)}+\sum_i{\lambda_i\Big(\sum_j q_{(i,j)}-1\Big)^2} \nonumber\\
&+&\sum_j{\mu_j \Big(\sum_i d_i q_{(i,j)}+\langle\bf{2},\bf{a}_j\rangle}-v_j \Big)^2 \nonumber\\
&+& \sum_j\alpha_j\Big(\tilde{q}_{(j)}-\sum_jq_{(i,j)}-\sum_jb_{(i,j)}\Big)^2.
\end{eqnarray}
Here $\tilde{q}_{(j)}$ denote the configurations of facilities at site $x_j$. The penalty scaled by $\alpha_j$ is only activated when $\tilde{q}_{(j)}=0$ and $\sum_jq_{(i,j)}\geq1$, i.e., customers are assigned to sites where no facility is built. The standard approach requires additional auxiliary qubits of $m\times n$ and enormous couplers for implementing the $\tilde{q}_{(j)}q_{(i,j)}$, $\tilde{q}_{(j)}b_{(i,j)}$, and $q_{(i,j)}b_{(i,j)}$ interaction, which is too complicated for minor embedding, which aims at find an effective Hamiltonian with the same low-energy subspace of Eq.~\eqref{eq:direct}. Meanwhile, diadiabatic effect can cause energy excitation during the annealing process, resulting in failure of preparing the ground state of the effective Hamiltonian. In this way, one may obtain quasi-optimal solutions or even solutions that violate the constraint conditions, if the energy gaps between low-energy states are not sufficiently large. This requires a systematic study of optimizing the quantum annealing of the problem Hamiltonian~\eqref{eq:direct} and its effective Hamiltonian for minor embedding, e.g., the customized annealing schedules and embedding algorithms, which goes beyond the scope of this work. Although we have compared the outcome of the hybrid algorithm with the global optimal result, a comparison of the performance between our hybrid classical-quantum algorithm and the standard classical approach~\eqref{eq:direct} would be relevant for a future research. We will undoubtedly consider this issue as a key milestone for a future followup project.

On the hardware side, the priority is to own a D-Wave quantum annealer instead of using a cloud quantum annealer, considering that the inner annealing time does not contribute a lot to the whole computation time. An advance could be controlling the quantum annealing process in the inner layer. Generally, the Hamiltonian of a quantum annealer $H=A(t)H_T+B(t)H_P$ is constrained by boundary conditions that $A(0)=1$, $A(t_f)=0$, $B(0)=0$ and $B(t_f)=1$, to start with initial tunneling Hamiltonian and result in the ground state (or low-energy state) of the final problem Hamiltonian. A customized quantum annealing process might shorten the annealing time while reducing the energy excitation to generate a better solution within less time. This annealing protocol could be given by control theory or other optimal methods, e.g., shortcut to adiabaticity in spin system~\cite{sta-spin-yu,sta-spin-zhang,sta-takahashi,sta-mori}, that controls the preparation and evolution of qubits in the quantum annealer. Quality of the solutions could also be improved by quantum annealer with more qubits, larger connectivity, and less noise, which will be released by D-Wave in mid-2020 named Pegasus~\cite{pegasus}. Alternative hardware will be coherent quantum annealer, which is still far from the practical application but could be built with current technologies while providing preliminary results~\cite{coherentDWave,LechnerSciAdv,Lechner}.

\section{Conclusion\label{sec:conclusion}}
We proposed a combined quantum annealing algorithm inspired in Ref.~\cite{CSA} to solve logistic network design problems, but which can also be applied to a large variety of optimization problems. The algorithm is tested with 12 NDP problems, and the results are in very good agreement with the already-known best solutions given by \textit{Lindo}. This research is another convincing evidence for the feasibility of applying quantum annealing for optimization problems, even when the quantum devices are limited by the number of qubits, the connectivity, and the noise.

\section{Acknowledgments}
We acknowledge funding from projects QMiCS (820505) and OpenSuperQ (820363) of the EU Flagship on Quantum Technologies, Spanish Government PGC2018-095113-B-I00 (MCIU/ AEI/FEDER, UE), PID2019-104002GB-C21, PID2019-104002GB-C22 (MCIU/AEI/FEDER, UE), Basque Government IT986-16, NSFC (Grant No. 12075145), Shanghai Municipal Science and Technology Commission (Grant No. 2019SHZDZX01-ZX04, 18010500400, and 18ZR1415500), and the Shanghai Program for Eastern Scholar, as well as the and EU FET Open Grant Quromorphic. This work is supported by the U.S. Department of Energy, Office of Science, Office of Advanced Scientific Computing Research (ASCR) quantum algorithm teams program, under field work proposal number ERKJ333.

\appendix
\section{Simulated Annealing Algorithm for NDPs\label{app:SA}}
Here we introduce how to perform simulated annealing algorithm for solving NDPs. We keep the same cooling schedule and iteration number at each cooling step for a fair comparison. The initial state is an allocation of customers that randomly distributed to different sites and its according configuration of facilities. If a site is visited by at least once, a facility should be built on this site. The neighboring function will be, select an arbitrary customer among all $n$ customers, and randomly assign the customer to a site. With the neighboring function and definition of state, one can evaluate the cost difference, updating the solution by Metropolis acceptance criterion as we do in outer layer of combined quantum annealing algorithm.


\begin{thebibliography}{99}

\bibitem{ballou} R. H. Ballou, "Logistics network design: modeling and informational considerations", The International Journal of Logistics Management {\bf 6}, 39 (1995).

\bibitem{sa} S. Kirkpatrick, C. D. Gelatt, and M. P. Vecchi, "Optimization by simulated annealing", Science {\bf 220}, 571 (1983).

\bibitem{tabu1} F. Glover, "Tabu search: part I", ORSA Journal on computing {\bf 1}, 190 (1989).

\bibitem{tabu2} F. Glover, "Tabu search: part II", ORSA Journal on computing {\bf 2}, 4 (1990).

\bibitem{ga} D. E. Goldberg and J. H. Holland, "Genetic algorithms and machine learning", Machine learning {\bf 3}, 95 (1988).

\bibitem{aco1} M. Dorigo, "Optimization, learning and natural algorithms" (PhD Dissertation, Politecnico di Milano, 1992).

\bibitem{aco2} M. Dorigo, V. Maniezzo, and A. Colorni, "Ant system: optimization by a colony of cooperating agents", IEEE Transactions on Systems, man, and cybernetics, Part B: Cybernetics {\bf 26}, 29 (1996).

\bibitem{ann} J. M. Zurada, {\it Introduction to artificial neural systems} (St. Paul: West publishing company, 1992).

\bibitem{prelim1} V. Jayaraman and A. Ross, "A simulated annealing methodology to distribution network design and management", European Journal of Operational Research {\bf 144}, 629 (2003).

\bibitem{prelim2} D. Ghosh, "Neighborhood search heuristics for the uncapacitated facility location problem", European Journal of Operational Research {\bf 150}, 150 (2003).

\bibitem{prelim3} M. Gen and A. Syarif, "Hybrid genetic algorithm for multi-time period production/distribution planning", Computers \& Industrial Engineering {\bf 48}, 799 (2005).

\bibitem{prelim4} M. Sun, Computers, "Solving the uncapacitated facility location problem using tabu search", \& Operations Research {\bf 33}, 2563 (2006).

\bibitem{annealing1} A. B. Finnila, M. A. Gomez, C. Sebenik, C. Stenson, and J. D. Doll, "Quantum annealing: A new method for minimizing multidimensional functions", Chemical Physical Letters  {\bf 219}, 343 (1994).

\bibitem{annealing2} A. Das and B. K. Chakrabarti, "Colloquium: Quantum annealing and analog quantum computation", Reviews of Modern Physics {\bf 80}, 1061 (2008).

\bibitem{nielsenchuang} M. A. Nielsen and I. L. Chuang, {\it Quantum Computation and Quantum Information} (Cambridge University Press, Cambridge, UK, 2000).

\bibitem{cmp-spinglass-theory} G. E. Santoro, R. Marto\v{n}\'ak, E. Tosatti, and R. Car, "Theory of quantum annealing of an Ising spin glass", Science {\bf 295}, 5564 (2002).

\bibitem{cmp-manufactured} M. W. Johnson, M. H. S. Amin, S. Gildert, T. Lanting, F. Hamze, N. Dickson, R. Harris, A. J. Berkley, J. Johansson, P. Bunyk, E. M. Chapple, C. Enderud, J. P. Hilton, K. Karimi, E. Ladizinsky, N. Ladizinsky, T. Oh, I. Perminov, C. Rich, M. C. Thom, E. Tolkacheva, C. J. S. Truncik, S. Uchaikin, J. Wang, B. Wilson, and G. Rose, "Quantum annealing with manufactured spins", Nature {\bf 473}, 104 (2011).

\bibitem{cmp-traverse} T. Kadowaki and H. Nishimori, "Quantum annealing in the transverse Ising model", Phys. Rev. E {\bf 58}, 5355 (1998).

\bibitem{cmp-disordered} J. Brooke, D. Bitko, and G. Aeppli, "Quantum annealing of a disordered magnet", Science {\bf 284}, 779 (1999).

\bibitem{cmp-experiment} R. Harris, Y. Sato, A. J. Berkley, M. Reis, F. Altomare, M. H. Amin, K. Boothby, P. Bunyk, C. Deng, C. Enderud, S. Huang, E. Hoskinson, M. W. Johnson, E. Ladizinsky, N. Ladizinsky, T. Lanting, R. Li, T. Medina, R. Molavi, R. Neufeld, T. Oh, I. Pavlov, I. Perminov, G. Poulin-Lamarre, C. Rich, A. Smirnov, L. Swenson, N. Tsai, M. Volkmann, J. Whittaker, and J. Yao, "Phase transitions in a programmable quantum spin glass simulator", Science {\bf 361}, 6398 (2018).

\bibitem{volkswagen} F. Neukart, G. Compostella, C. Seidel, D. Dollen, S. Yarkoni, and B. Parney, "Traffic flow optimization using a quantum annealer", Frontiers in ICT {\bf 4}, 29 (2017).

\bibitem{FengHu} F. Hu, L. Lamata, M. Sanz, X. Chen, X.-Y. Chen, C. Wang, and E. Solano, "Quantum computing cryptography: Unveiling cryptographic Boolean functions with quantum annealing", Phys. Lett. A {\bf 384}, 126214 (2020).

\bibitem{FengHu2} F. Hu, L. Lamata, C. Wang, X. Chen, E. Solano, and M. Sanz, "Quantum Advantage in Cryptography with a Low-Connectivity Quantum Annealer", Phys. Rev. Applied {\bf 13}, 054062 (2020).

\bibitem{bio} A. Perdomo-Ortiz, N. Dickson, M. Drew-Brook, G. Rose, and A. Aspuru-Guzik, "Finding low-energy conformations of lattice protein models by quantum annealing", Scientific Reports, {\bf 2}, 571 (2012).

\bibitem{quantfin1} G. Rosenberg, P. Haghnegahdar, P. Goddard, P. Carr, K. Wu, and M. L. de Prado, "Solving the optimal trading trajectory problem using a quantum annealer", IEEE Journal of Selected Topics in Signal Processing {\bf 10}, 1053 (2016).

\bibitem{quantfin2} R. Or\'us, S. Mugel, and E. Lizaso, "Quantum computing for finance: overview and prospects", Reviews in Physics {\bf 4}, 100028 (2019).

\bibitem{quantfin3} Y. Ding, L. Lamata, J. D. Mart\'in-Guerrero, E. Lizaso, S. Mugel, X. Chen, R. Or\`us, E. Solano, and M. Sanz, "Towards Prediction of Financial Crashes with a D-Wave Quantum Computer", arXiv:1904.05808 (2019)

\bibitem{FGGS00} E. Farhi, J. Goldstone, S. Gutmann, and M. Sipser, "Quantum Computation by Adiabatic Evolution", arxiv: quant-ph/0001106 (2000).

\bibitem{AE99} J. E. Avron and A. Elgart, "Adiabatic Theorem without a Gap Condition", Communications in Mathematical Physics {\bf 203}, 445 (1999).

\bibitem{CSA} J. Qin and L. X. Miao, "Combined simulated annealing algorithm for logistics network design problem", International Workshop on Intelligent Systems and Applications, IEEE, (2009).

\bibitem{OR-Library} J. Beasley, see as \url{http://people.brunel.ac.uk/~mastjjb/jeb/orl-ib/files/}

\bibitem{qbsolv} See, for example: \url{https://github.com/dwavesystems/qbsolv}

\bibitem{hajek} B. Hajek, "Cooling schedules for optimal annealing", Mathematics of Operations Research {\bf 13}, 311 (1988).

\bibitem{walid} W. Ben-Ameur, "Computing the initial temperature of simulated annealing", Computational Optimization and Applications {\bf 29}, 369 (2004).

\bibitem{improving-qbsolv} S. Okada, M. Ohzeki, T. Terabe, and S. Taguchi, "Improving solutions by embedding larger subproblems in a D-Wave quantum annealer", Scientific Reports {\bf 9}, 2098 (2019).

\bibitem{hua} A. A. Abbott, C. S. Calude, M. J. Dinneen, and R. Hua, "A hybrid quantum-classical paradigm to mitigate embedding costs in quantum annealing", International Journal of Quantum Information {\bf 17}, 1950042 (2019).

\bibitem{sta-spin-yu} X. T. Yu, Q. Zhang, Y. Ban, and X. Chen, "Fast and robust control of two interacting spins", Physical Review A {\bf 97}, 062317 (2018).

\bibitem{sta-spin-zhang} Q. Zhang, X. Chen, and D. Gu\'ery-Odelin, "Reverse engineering protocols for controlling spin dynamics", Scientific Reports {\bf 7}, 15814 (2017).

\bibitem{sta-takahashi} K. Takahashi, "Shortcuts to adiabaticity for quantum annealing", Physical Review A {\bf 95}, 012309 (2017).

\bibitem{sta-mori} T. Hatomura and T. Mori, "Shortcuts to adiabatic classical spin dynamics mimicking quantum annealing", Physical Review E {\bf 98}, 032136 (2018).

\bibitem{pegasus} See, for example: \url{www.dwavesys.com/press-releases/d-wave-previews-next-generation-quantum} \url{-computing-platform}

\bibitem{coherentDWave} I. Ozfidan, C. Deng, A. Y. Smirnov, T. Lanting, R. Harris, L. Swenson, J. Whittaker, F. Altomare, M. Babcock, C. Baron, A.J. Berkley, K. Boothby, H. Christiani, P. Bunyk, C. Enderud, B. Evert, M. Hager, A. Hajda, J. Hilton, S. Huang, E. Hoskinson, M.W. Johnson, K. Jooya, E. Ladizinsky, N. Ladizinsky, R. Li, A. MacDonald, D. Marsden, G. Marsden, T. Medina, R. Molavi, R. Neufeld, M. Nissen, M. Norouzpour, T. Oh, I. Pavlov, I. Perminov, G. Poulin-Lamarre, M. Reis, T. Prescott, C. Rich, Y. Sato, G. Sterling, N. Tsai, M. Volkmann, W. Wilkinson, J. Yao, and M. H. Amin, "Demonstration of nonstoquastic Hamiltonian in coupled superconducting flux qubits", arXiv:1903.06139

\bibitem{LechnerSciAdv} W. Lechner, P. Hauke, and P. Zoller, "A quantum annealing architecture with all-to-all connectivity from local interactions", Science Advances {\bf 1}, e1500838 (2015).

\bibitem{Lechner} P. Hauke, H. G. Katzgraber, W. Lechner, H. Nishimori, and W. D. Oliver, "Perspectives of quantum annealing: Methods and implementations", arXiv:1903.06559
\end{thebibliography}
\end{document}